# Beyond broad strokes: sociocultural insights from the study of ancient genomes

Fernando Racimo[1], Martin Sikora[1], Hannes Schroeder[1], Carles Lalueza-Fox[2]

1. Globe Institute, Faculty of Health and Medical Sciences, University of Copenhagen, Copenhagen, Denmark
2. Institute of Evolutionary Biology, CSIC-Universitat Pompeu Fabra, 08003 Barcelona, Spain

**Abstract**

The amount of sequence data obtained from ancient samples has dramatically expanded in the last decade, and so have the types of questions that can now be addressed using ancient DNA. In the field of human history, while ancient DNA has provided answers to long-standing debates about major movements of people, it has also recently begun to inform on other important facets of the human experience. The field is now moving from not only focusing on large-scale supra-regional studies to also taking a more local perspective, shedding light on socioeconomic processes, inheritance rules, marriage practices and technological diffusion. In this review, we summarize recent studies showcasing these types of insights, focusing on the methods used to infer sociocultural aspects of human behaviour. This work often involves working across disciplines that have, until recently, evolved in separation. We argue that multidisciplinary dialogue is crucial for a more integrated and richer reconstruction of human history, as it can yield extraordinary insights about past societies, reproductive behaviours and even lifestyle habits that would not have been possible to obtain otherwise.

**Introduction**

In recent years, the field of archaeogenomics has shed new light on the timing and composition of ancient migrations, and how they shaped present-day human diversity[1–3]. Thanks to explosive improvements in methods for extracting and sequencing ancient DNA, the number of available ancient genomes has jumped from less than 5 just a decade ago to over three thousand at the moment of writing. Additional improvements in bioinformatics and population genetic inference have also served to extract invaluable information from these genomes, including patterns of population growth and contraction, interbreeding between distantly related groups and evidence for natural selection operating on phenotypically important loci[4,5].

Ancient DNA has also served to inform long-standing debates in archaeology, particularly about the role of "demic"[6–8] vs. "cultural" diffusion[9] in the spread of technologies. Since the 1970s, the emergence of molecular studies and statistical migration models placed new emphasis on population movements as the driver for technological expansions[10], like the spread of agriculture[11]. Recent archaeogenetic analyses have tipped the scales towards a view of the past largely influenced by migrations. Rather than complete population replacements, however, the emerging trends point towards population admixture as the process by which cultures and technologies tend to spread into new regions[12]. In Europe, for example, the retrieval and analysis of genomes from the past 15,000 years indicate that the advent of agriculture was associated with admixture between farming and hunter-gathering populations at a continental scale[13,14]. A subsequent migration of nomads from the Eurasian steppe left an equally important footprint in the genomes of present-day people, and may have introduced proto-Indo-European languages into Europe[15]. Similar processes have been inferred in the transition to agriculture in other regions of the world – including Central-South Asia[16] and Southeast Asia[17,18] – where present-day individuals bear diverse ancestries from both original hunter-gatherer populations and from subsequent migrations.

As sample sizes have increased, more recent studies have shifted their focus towards understanding more micro-level phenomena (Figure 1). These range from marriage practices and family burial customs to social organization and even ancient conflicts. This shift has been aided by methodological developments, which have pushed down the cost of sequencing ancient genomes[19–22], and enabled the bulk sequencing of dozens or even hundreds of individuals from the same region or locality. Researchers are no longer forced to pick particularly valuable or well preserved samples, but can obtain genomic data from entire temporal transects of a region, cemetery or depositional site[23,24].

The shift has also been catalyzed by a dialogue with other disciplines, which can provide invaluable historical and biological context to archaeogenomic findings. Major ancient DNA consortia across the world now include experts in archaeology, linguistics, history, ecology, and radiometric dating. Not only do they provide invaluable assistance to geneticists in understanding the importance of a particular finding, but also help them frame questions and posit hypotheses that can best serve to inform long-standing debates. Moreover, they can bridge the gap between "genome" and "person" – and remind them that behind the sequence of As, Cs, Ts and Gs lies a human being with potentially important genealogical or cultural ties to the present-day inhabitants of a region.

Working in close collaboration with other natural and social scientists of the past, ancient DNA researchers are now in a position to provide a more complete view of long-gone societies, and to test previously unverifiable hypotheses about local changes in cultural practices over time. In this review, we explore some of these insights, and provide recommendations for building a more integrative approach to the study of the past.

**Techno-cultural development and diffusion**

Ancient DNA provides a unique window into the process of technology transfer and development. They can allow us to determine whether these transfers tend to occur via population replacement, intermixing between populations or without either of these, i.e. do genes and technologies move together? A canonical example of technological diffusion is the spread of farming over the European continent during the Neolithic period. A plethora of radiocarbon-dated sites showing early evidence of farming have allowed archaeologists to model how this cultural innovation spread throughout Europe in the Neolithic. Consistently across studies, researchers have observed that the expansion of farming was broadly two-pronged, consisting of two cultural movements which can be broadly associated with the Impressa / Cardial Ware and Linear Pottery archaeological horizons[25–29]. Now, the new abundance of radiocarbon-dated ancient genomes allows us to trace how genetic ancestry spread as a consequence of this cultural transformation[14,30]. Indeed, the "genetic expansion" of ancestry during the Neolithic appears to have a similar shape to that evinced by archaeology[31]. This supports the idea that movements of people originating from Anatolia and the Middle East brought about this transformation throughout Europe (Figure 2).

Ancient genomic studies can also inform on the biological consequences of technological innovations. For example, they can inform on how new selective pressures affected the frequency of particular genetic variants involved in diet and metabolism. Buckley et al. and Ye et al.[32,33] showed that a cluster of genes coding for fatty acid desaturases have been under strong positive selection since the Bronze Age. They argued that this likely occurred because of a change to a diet rich in fatty acids derived from plant sources, after the Neolithic transition to agriculture. More recently, Mathieson and Mathieson[34] showed that the selective pressures on this gene cluster occurred several hundred years after the transition to agriculture. They argued that the shift in allele frequencies may have occurred due to more recent changes in diet or environment, or perhaps as a consequence of increased population sizes and, therefore, an increased efficiency of natural selection.

Studies of cultural transformations can also benefit from archaeogenomic research, as they can reveal whether the influx of new cultures was mediated by admixture[35]. An example of this is the spread of Bell Beaker pottery across western and central Europe between 2,750 and 2,500 BCE, and its subsequent disappearance around 2,000 BCE. A recent study showed that there are few genetic connections between individuals associated with this type of pottery in Iberia and central Europe, suggesting its continental dissemination was largely driven by cultural diffusion rather than migration. Its spread into the British Isles, however, was strongly driven by migration of peoples, and may have led to a replacement of a large portion of the British population in only a few hundred years[36].

A third important techno-cultural development informed by ancient genomics is horse breeding and herding. Archaeological evidence suggests that the Botai people – who lived in Kazakhstan between 3,500 and 3,000 BCE – were the earliest population to tame horses and breed them. It has been argued that they incorporated this practice via migration from the Yamnaya, another group of horse herders that likely introduced this practice into Europe via a large-scale migration in the Bronze Age[5,15]. A recent study, however, showed that the Botai and the Yamnaya do not share strong genetic affinities, and that the Botai are closely related to Siberian Paleolithic hunter-gatherer groups[37]. Furthermore, another study on ancient horse DNA indicates that the Botai horses are not particularly closely related to modern horse breeds[38]. Taken together, this suggests there may have been two independent horse domestication events, one in Western Eurasia and one in Central Eurasia[37]. On a more recent timescale, time-series genomic data from horses has revealed the existence of previously unknown domestic horse lineages in Iberia and Siberia during the early stages of domestication, which did not leave much descendancy in present-day populations[39]. This study also showed that domestication brought about a severe decline in genetic diversity, which is consistent with a strong reduction in horse breeding stock during the last few centuries.

**Kinship and social organization**

Because genomes contain information about genealogical relationships (Box 1) and inbreeding (Box 2, Figure 3), archaeogenetics can also help us better understand patterns of social organization in societies with little to no written records. A recent study based on Neolithic genomes from Western Europe focused on human remains located in megalithic tomb burials – which became common around 4,500 BCE in the Atlantic coast and in the British Isles. The authors found a significant excess of males relative to females in these burials, and found the same Y-chromosome haplotypes in burials re-occurring across different time periods. They thus suggested that these burials were likely linked to stable groups that followed a patrilineal form of social organization, and that the role of the burials was to harbor the remains of those related to particular paternal lineages[40].

In a different study, Schroeder et al.[41] sequenced the genomes of 15 individuals from a Late Neolithic mass grave in Poland belonging to the so-called Globular Amphora Culture (3,300–2,700 BCE). The grave contained the remains of three generations of men, women and children, all of whom had been brutally killed by blows to the head. The genetic analyses revealed that the individuals were part of the same extended family group, as they were almost all related to each other through various first- (parent-offspring or siblings) and second-degree relationships (aunts, uncles, half-siblings, etc.). However, while the men all appeared to be closely related (through paternal lines of descent), the women were much more genetically diverse, suggesting that this Late Neolithic community may also have been organized around patrilocal or virilocal residence patterns. Overall, these findings fit with previous studies[42,43], suggesting that patrilocality and female exogamy may have been the dominant form of social organisation in Europe at the end of the Neolithic.

In other cases, ancient DNA studies have demonstrated that certain cultures or regions preserved matrilineal continuity for long periods of time. Kennett et al.[44] retrieved aDNA from 9 individuals buried in an elite crypt in Chaco Canyon over a period of more than 300 years. They found all individuals had identical mitochondrial genomes and some had genealogical affinities consistent with matrilineal continuity: mother-daughter and grandmother-granddaughter. Margaryan et al.[45] recovered mitochondrial genomes from the South Caucasus from between 300 and 7,811 years ago. They used population genetic simulations to show that a model of long-term continuity of the maternal gene pool was most consistent with the patterns of mitochondrial variation, in spite of well-documented cultural shifts that occurred in the region over that time span.

**Social status and inequality**

Generally, archaeological data (like information about grave goods and burial customs) are essential for inferring patterns of social status and inequality. These data can be analyzed in conjunction with genetic data in order to discern inheritance rules and patterns of social organization. In a groundbreaking study, Mittnik et al.[46] recently combined ancient genomes, isotope analyses, and archaeological information from grave goods found in the Lech River valley in southern Germany during the Late Neolithic to Middle Bronze Age (2,800-1300 BCE). They demonstrated that individual farmsteads in this region generally consisted of a high-status core family and unrelated low-status individuals, who lived together under one roof – a form of social organization that persisted over centuries[46].

Ancient genomes can also reveal whether social or religious status was associated with populations originating from distant or nearby lands. For instance, a study on Longobard burials from two necropoles in northern Italy and Hungary, dating from the 5th-7th centuries CE, revealed that richly-endowed burials showed a higher proportion of central European ancestry than other local burials. The presence of this ancestry is consistent with a migration from Pannonia into Northern Italy after the fall of the Roman Empire, and suggests that, at the time, individuals with high social status were likely members of this recent migration into the region[47].

In another recent study, Narasimhan et al. collected hundreds of ancient genomes from South and Central Asia, and showed that a Pontic Steppe pastoralist migration penetrated into South Asia during the second millennium BCE[16]. This migration may have brought Indo-European languages into the region, similarly to what is hypothesized to have occurred in Western Eurasia[15]. Different present-day South Asian groups possess genetic ancestry originating from this migration in different proportions. Intriguingly, those groups that traditionally consider themselves as having priestly status, like the Brahmins, have significantly higher proportions of this ancestry than other South Asian groups. The authors suggest that an explanation for this pattern could be that the descendants of the Steppe migration were the traditional custodians of literature written in early Sanskrit. Perhaps, the extreme endogamy characteristic of this region led these groups to remain in relative isolation from other South Asian groups over thousands of years, thus leading to the present-day correlation between Steppe ancestry and priestly status[16].

**Sex biases in migration**

As genetic females carry two copies of the X chromosome while males only carry one, this means that lineages in the X chromosome tend to exist inside the female germline twice as often as in the male germline. This allows geneticists to explore if a migration is male-driven by exploring the fraction of a specific ancestry in the X chromosomes as compared to the autosomes in the same individual. The combination of these analyses with analyses based on uniparental markers can provide evidence for gender-biased migrations. Genetic evidence suggests that certain migratory movements were extremely sex-biased[48] (although see also refs. [49] and [50]), with incoming groups mainly formed by males organized in patrilineal clans that admixed with local women, in processes that resulted in language substitutions and also in the emergence of social elites.

Olalde et al.[51] detected a previously unrecognized genomic turnover during a 400-year period in the Iberian Bronze Age. The scale of this replacement is large (about 40% of the total ancestry was replaced), but is also remarkably gender-biased: about 100% of all Y-chromosomes were replaced. The Iberian X-chromosomes showed half the steppe ancestry ratio detected in the autosomes of the same individuals (17.3% vs 38.9%), thus indicating that this process was mainly driven by incoming males. Another instance of male-biased replacement occurred in South Asia, where genetic material from Bronze Age Steppe pastoralists was introduced mostly via males[16]. In Estonia, the spread of Steppe ancestry via the Corded Ware Culture expansion also appears to have been male-biased, although this expansion additionally carried early farmer ancestry with a female bias[52]. Focusing on more recent samples, Sandoval-Velasco et al.[53] detected a strong male sex bias in a burial population of liberated Africans who died on the island of St Helena, which reflects broader patterns in the latter phases of the transatlantic slave trade that are well attested historically.

**Individual journeys and mobility**

Archaeogenetic studies are also uncovering personal stories of people who in some cases had remarkable journeys. We may never know the specific details of their travels, but traces of their movement can be evinced from unlikely patterns of ancestry or relatedness (Box 1). For example, Olalde et al.[51] found that a 4,400-year-old individual found buried near Madrid had North African ancestry, suggesting he or his direct ancestors may have been recent migrants into the Iberian peninsula[51]. Moreover, the same study revealed that a woman from a V-VIth century CE Visigothic site in Girona (Northeast Spain) had clear affinities to Eastern Europeans, and could therefore have travelled from there or have relatives there, perhaps as a consequence of the recent Goth invasion into the Roman Empire.

In other cases, a genomic perspective can also inform us about the nature of voyages or pilgrimages, in cases where no other archaeological artefacts remain. A prime example of this is Roopkund Lake – a Himalayan site where hundreds of skeletal remains of unknown provenance have been found at over 5,000 meters above sea level. Harney et al.[54] performed genomic analyses on 23 individuals from this site, and found that just about half of the remains were from individuals with South Asian ancestry, dated to ~800 CE. The authors concluded that they may have died while performing the Nanda Devi Raj Jat pilgrimage, suggesting that this religious practice may have existed in some form at this time in the past. Most of the other remains came from unrelated individuals with eastern Mediterranean ancestry, and were dated to the 18th or 19th centuries. The reason why the latter group was there is unclear, but one possibility is that they were born in or near Greece, and eventually traveled to the Himalayas – perhaps as part of a different pilgrimage.

On a larger scale, ancient genomes can provide important insights into patterns of mobility during different periods of history and prehistory. Loog et al.[55] developed a statistical method based on allele frequency differentiation at individual loci in space and time, in order to test how much differentiation patterns are consistent with periods of high or low mobility. The authors used this method to show that mobility among Holocene farmers in Europe was significantly higher than among European hunter–gatherers both pre- and postdating the Last Glacial Maximum (see Box 3 and Figure 4).

Patterns of mobility may also be reflected in long-distance genetic kin relations. Margaryan et al.[56] found that several Viking individuals buried hundreds of kilometers apart were actually close relatives. The most surprising of these relations is a pair of 2nd degree male relatives (half-brothers, grandson-grandfather or nephew-uncle) that were buried on either side of the North Sea: one in Funen, Denmark and the other in Oxford, UK. Another pair of 3rd or 4th degree relatives (for example, cousins) was found in two Swedish Viking burials over 300 kilometers apart: in Skämsta and Öland. This suggests a high degree of mobility in Northern Europe during this period, at least within the social classes associated with Viking ornaments. Furthermore, the same study showed that the only Viking expedition with distinct archaeological traces, in Salme, Estonia, had a highly homogeneous ancestry profile, and contained several closely-related individuals, including four brothers. This suggests that elite Viking expeditions may have been carried out by individuals with the same ancestral origins, perhaps from the same village or region[56].

**Objects, identity and microhistories**

Retrieving ancient DNA from objects of common use can provide important insights about identity, life habits and behavior in the past. In two separate studies, Jensen et al.[57] and Kashuba et al.[58] recovered ancient DNA fragments from thousands-of-years-old pieces of "chewing gum". Jensen et al.[57] managed to reconstruct a complete ancient human genome from these fragments, which revealed aspects of the person who chewed the gum: a female individual with affinities to western European hunter-gatherers. The sample also contained fragments of microbial DNA, which provided information on her health status (she was infected with Epstein-Barr virus), and even faunal DNA fragments that probably derived from her diet.

In another object-based study, Schlabitsky et al.[59] extracted DNA from a nineteenth-century tobacco pipestem found on a former plantation site in Maryland, USA. Sequencing the DNA revealed that the pipe was used by a female individual of African ancestry. Closer analysis revealed that she had affinities to the Mende in Sierra Leone, suggesting that she or her ancestors may have originated there.

Sequencing DNA from personal objects like pipestems or gum has great potential for developing the concept of 'microhistories', which involves using a single artefact, life or incident as points of departure from which broader historical narratives can be developed[60]. Through ancient DNA analysis, scientists are now able to recover aspects of peoples' lives that were once thought unknowable. This opens up new avenues of inquiry that focus on the individual and their place in history[61].

**Conclusions and future perspectives**

Advances in archeometry, linguistics, bioinformatics, genomics and proteomics have revolutionized our understanding of history and prehistory: the study of the past is now blurring the lines between science and the humanities. It is increasingly clear that the reconstruction of human migrations and their social consequences will be a complex enterprise that can only be addressed by multidisciplinary teams[62]. Rigorous statistical methods and powerful computational techniques must be put into play in order to understand the massive amounts of data being produced by all these disciplines.

A keen awareness of these methodologies – as well as the possibilities and limitations of each field – will require new levels of interdisciplinary communication and statistical rigour. One could, for example, venture that statistical technologies used for inferring past divergence and admixture events from ancient genomes[63,64] could be translated into the linguistic realm, for inferring the divergence and admixture of ancient languages. Alternatively, agent-based models commonly used for reconstructing the spread of technologies or cultures in archaeology[29] could be used for modelling the spread of particular genetic ancestries over time and space. This, in turn, could help us distinguish which processes in history are best described as purely "demic," purely "cultural" or a combination of both[28,29]. One way to accelerate the process of integration with other disciplines is by establishing new archaeological science programs that train scientists jointly in all these disciplines. Under this framework, future MSc and PhD programs should not be defined by the methods that archeo-scientists can use, but by the questions they aim to answer.

As the number of sequenced ancient genomes scale up to several thousand, ancient DNA researchers should also keep in mind a number of ethical issues that are integral to their work. At the moment, extracting DNA from fossil material implies the partial destruction of said material, which may have important morphological, contextual, cultural or historical value, beyond the value provided by the acquisition of genetic information. Fox and Hawks[65] recently emphasized that researchers should keep careful records of sampled material and the results of sampling, and encourage accountability for both negative and positive results. They also highlighted the importance of formally engaging with stakeholders, including indigenous communities or close relatives that may have cultural or emotional connections with the material. Continuous discussions with local researchers – like anthropologists, social scientists and ecologists, who have long-standing ties to these communities – will facilitate this process and ensure a diverse set of voices can be heard.

We believe that ancient genomics should no longer be considered as separate realms from archaeology, history, anthropology or linguistics, but as another set of tools in the ever expanding methodological kit used for reconstructing ancient cultures – a toolkit that already includes decades-old scientific techniques, such as radiocarbon dating or isotope analysis. Just as our history seems to be characterized by extensive admixture among populations, it may be time to recognize that "admixture" among fields is the most optimal way forward for understanding our past.

**Acknowledgements**

We would like to thank Marc Vander Linden and Fabio Silva for providing raw data for reproducing agricultural spread patterns featured in a previous paper by them. FR was funded by a Villum Young Investigator award (project no. 000253000). HS was supported by HERA (Humanities in the European Research Area) through the joint research programme "Uses of the Past" and the European Union's Horizon 2020 research and innovation programme under grant agreement no. 649307. CL-F was supported by a grant from Obra Social "La Caixa" and FEDER-MINECO (PGC2018-095931-B-100) of Spain.

----
*Box 1 - Inferring relatedness using ancient genomes*

Close relatedness between individuals can be described via Cotterman's three k-coefficients[66]: $k_0$, $k_1$ and $k_2$. These coefficients characterize the amount of sharing of alleles due to shared ancestry between two non-inbred diploid genomes that are related in a particular way. $k_0$ is the probability that two individuals have 0 alleles that are identical by descent (IBD) at a random site in the genome, $k_1$ is the probability that they have 1 allele that is IBD at the site, and $k_2$ is the probability that both alleles at the site are IBD. Different types of genealogical relationships lead to different expectations for these coefficients:

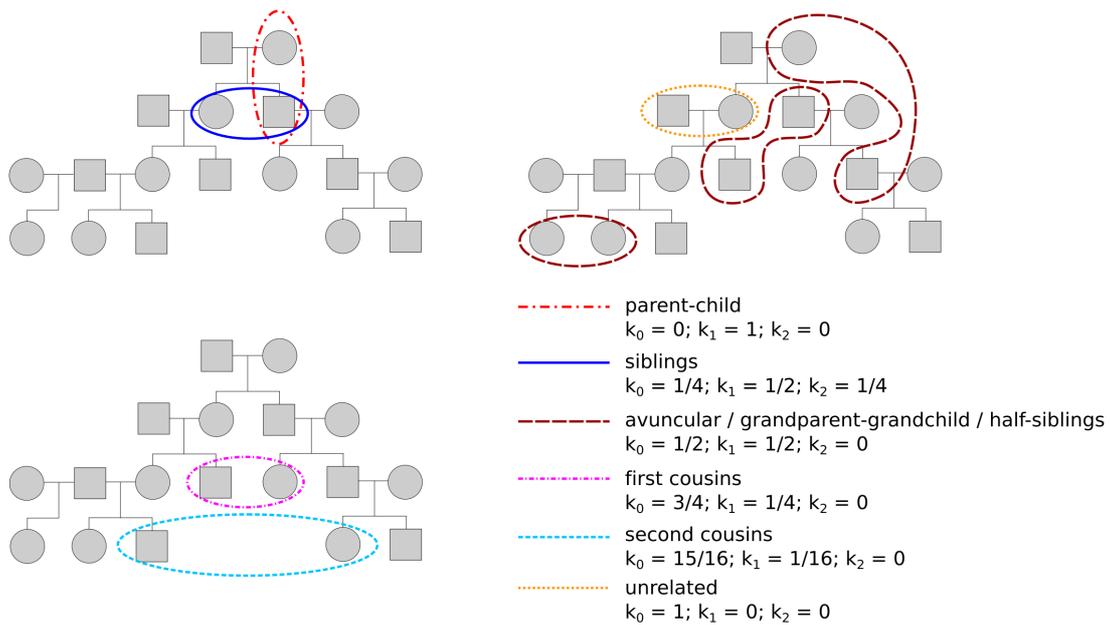

- ─·─·─ parent-child
  $k_0 = 0$; $k_1 = 1$; $k_2 = 0$
- ───── siblings
  $k_0 = 1/4$; $k_1 = 1/2$; $k_2 = 1/4$
- ─ ─ ─ avuncular / grandparent-grandchild / half-siblings
  $k_0 = 1/2$; $k_1 = 1/2$; $k_2 = 0$
- ─·─·─ first cousins
  $k_0 = 3/4$; $k_1 = 1/4$; $k_2 = 0$
- ─ ─ ─ second cousins
  $k_0 = 15/16$; $k_1 = 1/16$; $k_2 = 0$
- ·········· unrelated
  $k_0 = 1$; $k_1 = 0$; $k_2 = 0$

The problem is that these IBD probabilities are not known, and must be estimated from comparisons between diploid genomic sequences. The standard approach for doing so relies on the fraction of sites or haplotypes in the genome that are identical-by-state (IBS), i.e. loci where two individuals share the same allele. Researchers usually aim to maximize the likelihood of the IBS patterns observed in the data, given these k-coefficients, often assuming independence among loci:

$$L[\,data\,|\,k_0, k_1, k_2\,] = \prod_{i=1}^{N} \sum_{j=0}^{2} P[\,G_i\,|\,n = j\,]\,k_j$$

Here, N is the number of sites analyzed, G represents genotype data and n is the (unknown) number of alleles that are IBD at a particular locus[67,68]. Crucially, the conditional probability $P[\,G_i\,|\,n = j\,]$ depends on the allele frequencies of each locus in the population to which the individuals belong. There are numerous implementations and variations of this general approach[67,69–72]. However, few of these are particularly useful when working with ancient genomes, which often have post-mortem damage and contamination, and tend to be sequenced at low coverage. Additionally, knowledge of population allele frequencies for ancient populations is often missing.

In recent years, several researchers have tried to address some of these issues. Korneliussen and Moltke[73] developed the first tool to estimate relatedness that did not require called genotypes, and could be used with genotype likelihoods obtained from low-coverage genomes. Martin et al.[74] created a simulation-based method to estimate relatedness among ancient genomes using genetic distances, while accounting for sequencing error and contamination from present-day individuals. Theunert et al.[75] developed a maximum-likelihood method that can jointly estimate levels of contamination, sequencing error rates and pairwise relatedness coefficients from a set of ancient genomes, even if the samples are highly contaminated. Kuhn et al.[76] built a method particularly tailored for ancient genomes with low coverage, using pseudo-haploid random-read sampling – a standard practice in archaeogenomics. More recently, Waples et al.[77] developed a maximum-likelihood method for estimating relatedness using read data from low-coverage genomes, without the need for population allele frequency information.

----

----
*Box 2 - Estimating inbreeding from shared haplotypes*

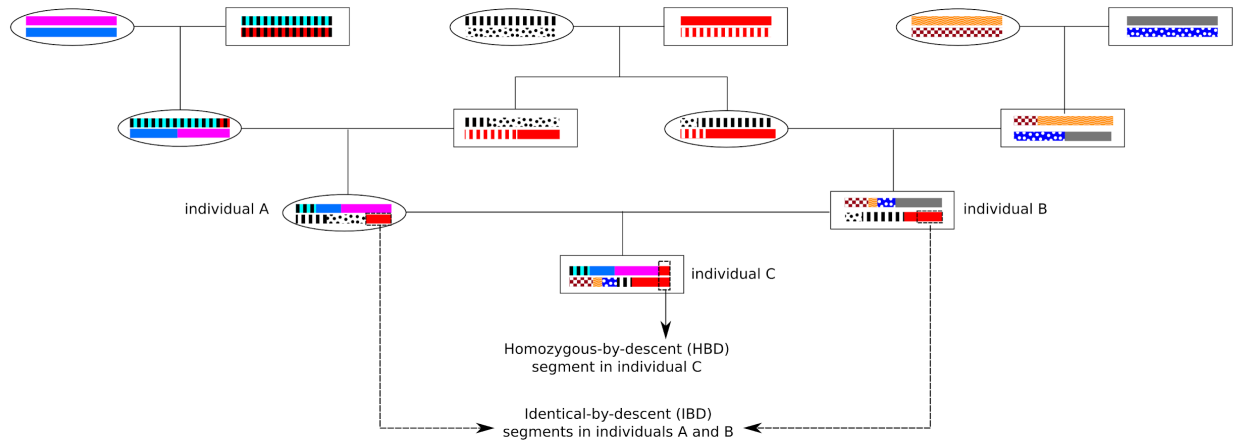

Identity-by-descent (IBD) is a fundamental concept in population genetics describing shared ancestry among genetic relatives. A pair of genomic segments or haplotypes is considered IBD if they were both inherited from a recently shared common ancestor. The amount of IBD sharing observed across the genomes of two individuals is therefore informative about their genetic relationship. For closely related individuals with very recent co-ancestry, the proportion of their genomes shared IBD is high, and is contained within long segments. With decreasing relatedness between individuals and increasing number of meioses separating them, their shared IBD proportion decreases exponentially, and IBD segments are broken into smaller chunks through recombination.

A pair of haplotypes shared IBD within the same individual is referred to as autozygosity, or homozygosity-by-descent (HBD). As two close genetic relatives share a large fraction of their genome in long IBD segments, their mating (inbreeding) will result in high amount of HBD in the resulting offspring. The distributions of the number and lengths of HBD segments observed within individuals in a population is thus informative about their demographic history. For example, a past population bottleneck results in a large number of short HBD segments in the descendant individuals, due to many genomic loci inherited IBD from a small number of founding individuals. Offspring from recent consanguineous mating on the other hand is expected to result in fewer, but substantially longer HBD segments. Contrasting the fraction of the genome contained in long versus short HBD segments is therefore a straightforward way to infer the impact of recent inbreeding in a population (Figure 4). Quantifying the fraction of HBD within an individual also provides a direct estimator for their inbreeding coefficient *F*, the probability that two alleles at a genomic locus are inherited IBD. For example, the inbreeding coefficient, or equivalently the expected fraction of the genome contained in HBD segments, for offspring from first cousin marriages is $F=0.0625$ [78].

Detection of HBD from genomic data is achieved by detecting long, continuous stretches of homozygous genotypes, termed runs of homozygosity (ROH). A range of methods for ROH detection exist[71,72,79–83], but their reliance on accurate diploid genotypes makes their application to ancient human genomes only feasible for individuals with high genomic coverage. More recent approaches based on genotype likelihoods circumvent this issue, but require population allele frequencies which are generally not available for ancient populations[84]. Despite these limitations, some recent studies have provided intriguing insights into reproductive behavior and social organizations of early humans. Ancient genomes from Neanderthals and Denisovans showed low heterozygosity compared to modern humans, indicating that organization in small isolated populations may have been predominant for archaic hominins[85–88]. Furthermore, a Neanderthal individual from the Denisova cave in Siberia was found to carry a large fraction of HBD segments, many of them longer than 10 centiMorgans (cM). Using simulations of different inbreeding scenarios, Prufer et al[86] inferred that the individual was likely the offspring of parents as closely related as half siblings. These observations are in contrast to those obtained from ancient modern humans. ROH length distributions of Pleistocene hunter-gatherers are consistent with small effective population sizes, but show no evidence for recent consanguinity (Figure 3). In a study of four contemporaneous individuals from an Upper Paleolithic burial at Sunghir, Sikora et al[89] used the IBD sharing distributions to estimate their recent effective population size ($N_e$). Under an idealized Wright-Fisher population model, Palamara et al[90] derived an estimator of $N_e$ based on the lengths $l_i$ of IBD segments between $n$ haplotypes above a length threshold $u$ in a genome of length $\gamma$

$$\widetilde{N}_e = \frac{50(1 - \hat{p}_r + \sqrt{1 - \hat{p}_r})}{u\hat{p}_r}$$

with

$$\hat{p}_r = \frac{\sum_i l_i}{\left[\gamma \binom{n}{2}\right]}$$

Using HBD segments, the effective population sizes for Sunghir were estimated to be 200-500 individuals, suggesting cultural practices that emphasized exogamy and avoided recent inbreeding despite low population densities[89].

---

----
*Box 3 - Mobility estimation from ancient DNA*

Loog et al.[55] devised an estimator for the amount of mobility that existed among people in a region over different time periods. The estimator has good power to detect changes in this parameter over time, assuming densely sampled ancient DNA data from individuals in a region is available for the temporal transect of interest.

To assess how much mobility existed over a period of time, one can test to what extent patterns of genetic relatedness are best explained by physical distances, between individuals as opposed to temporal distance between individuals. In order to do so, one must first calculate a matrix G of geographic distances, a matrix T of temporal distances and a matrix M of genetic distances among all individuals for which there is available genetic data.

If genetic relatedness is primarily explained by geographic distances during a particular period of time, one would expect that period to be one of low mobility. If the opposite is true, one would expect that to be a period of high mobility. The problem is that one must find a way to assess how much weight should be placed on these two sets of variables. However, time and space are not measured on the same scales, i.e. it is non-trivial to find a way to compare 100 kilometers to 100 years. The authors solved this issue by devising a combined space-time distance matrix D, where an unknown scaling factor relates time to space, and each entry is equal to:

$$D_{ij} = \sqrt{G_{ij}^2 + (ST_{ij})^2}$$

Here, $D_{ij}$ is the space-time distance between individual i and individual j, $G_{ij}$ is the geographic distance and $T_{ij}$ is the temporal distance. A natural estimator for mobility is then equal to the particular value of the scaling factor S that maximizes the correlation between the genetic distance matrix and the space-time matrix.

Because S could potentially be infinite (in a scenario where genetic distances are entirely explained by temporal distances), the authors decided to take S to be the tangent of an angle α that can range from 0 to 90 degrees. This way, estimated values of α close to 90 degrees reflect periods of high mobility, while estimated values of α close to 0 degrees reflect periods of low mobility. The authors tested a range of values for α between 0 and 90 for various time periods of human European history, to find the particular angle (amount of mobility) that best correlated to the genetic distances between individuals in each period.

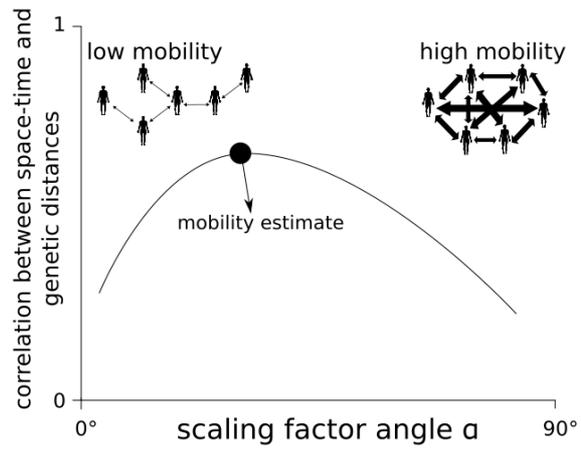

---

14/28

# Figures

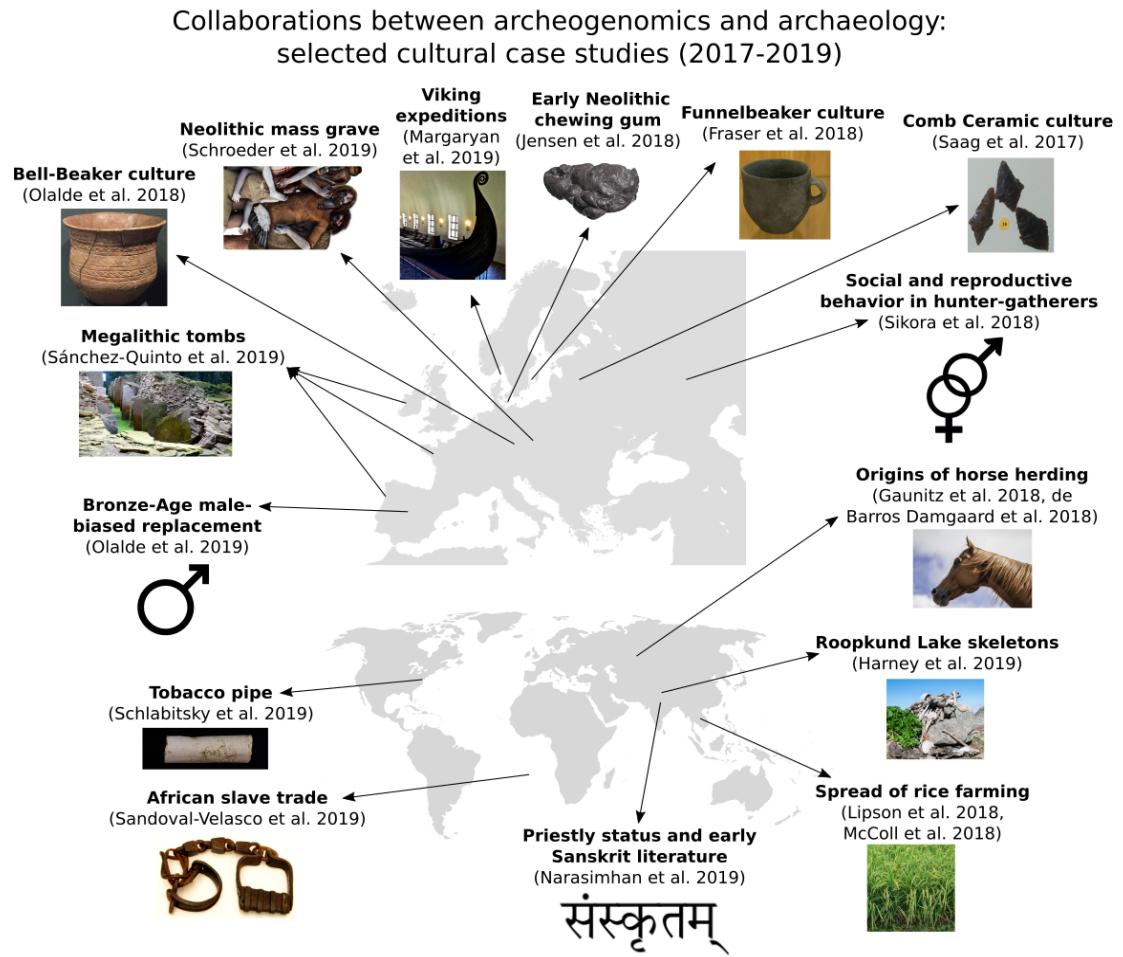

**Figure 1.** Case-studies: Schematic map of the world with arrows pointing to particular ancient genomic case studies informing on cultural processes of particular periods[17,36–38,40,41,51,52,54,56,59,89,91].
Image credits:
Rice: CC-BY - David Pursehouse https://www.flickr.com/photos/58246614@N00/2930951478
Horse: CC-BY - Beneharo Mesa - https://www.flickr.com/photos/135103535@N05/33427696864
Megalithic tomb: CC-BY -  John M. Schulze -
https://www.flickr.com/photos/13227526@N00/35020052064
Bell-beaker: CC-BY-SA - Zde -
https://commons.wikimedia.org/wiki/File:Bell_Beaker,_Copper_Age,_City_of_Prague_Museum,_175634.jpg
Funnelbeaker: CC-BY-SA - Zde -
https://commons.wikimedia.org/wiki/File:Funnel_Beaker_culture_pottery,_Museum_of_Kladno,_176030.jpg
Sanskrit script: CC-BY-SA - OldakQuill -
https://en.wikipedia.org/wiki/Sanskrit#/media/File:The_word_%E0%A4%B8%E0%A4%82%E0%A4%B8%E0%A5%8D%E0%A4%95%E0%A5%83%E0%A4%A4%E0%A4%AE%E0%A5%8D_(Sanskrit)_in_Sanskrit.svg
Slave shackle: CC-YB-SA - ZekethePhotographer - https://commons.wikimedia.org/wiki/File:Trans-Atlantic_Slave_Trade_Artifacts.png
Roopkund Lake: CC-BY-SA - Schwiki -
https://en.wikipedia.org/wiki/Roopkund#/media/File:Human_Skeletons_in_Roopkund_Lake.jpg
Viking ship: CC-BY - Larry Lamsa - https://www.flickr.com/photos/22191277@N03/22756327930

Comb Ceramic: Public Domain - Jānis U. - https://commons.wikimedia.org/wiki/Category:Comb_Ceramic_Culture#/media/File:Sarnates_apmetnes_depozits-2.jpg
Gum: courtesy Theis Trolle Zetner Jensen
Pipe stem: courtesy Julie Schablistky
Neolithic mass grave: courtesy Michał Podsiadło

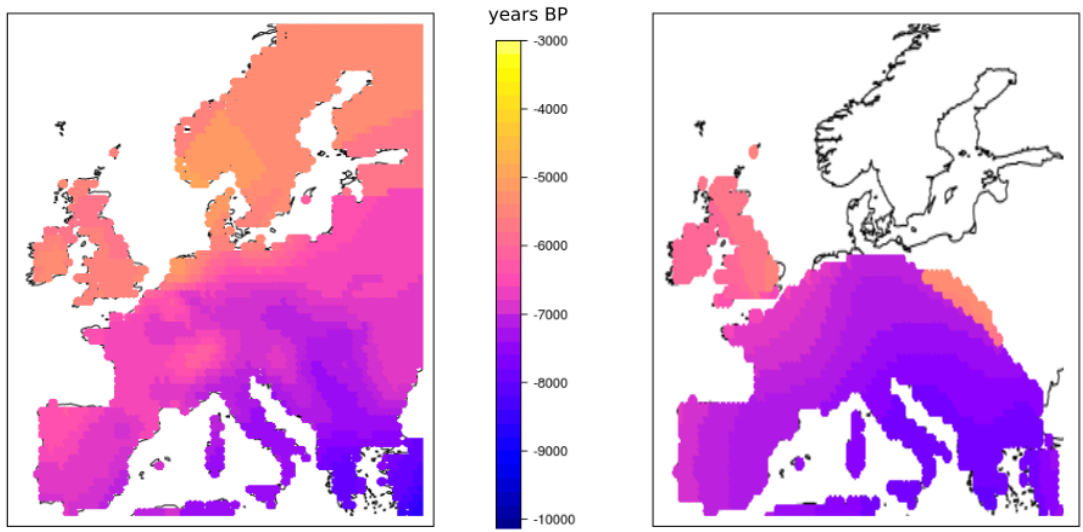

**Figure 2.** Spatio-temporally kriged first-arrival reconstructions for the spread of farming based on radiocarbon-dated archaeological sites, and for the spread of farming ancestry based on radiocarbon-dated ancient genomes. Grid data for building the left panel was obtained courtesy of Marc Vander Linden and Fabio Silva and is based on a figure from Linden and Silva[92]. The right panel was built after spatio-temporally kriging a collection of ancient genomes from Western Eurasian palaeogenomic studies mentioned in the main text, and recording the grid points at which one first encounters Anatolian farmer ancestry at a fraction higher than 75%[31]. White regions are regions where this ancestry did not reach levels above 75%.

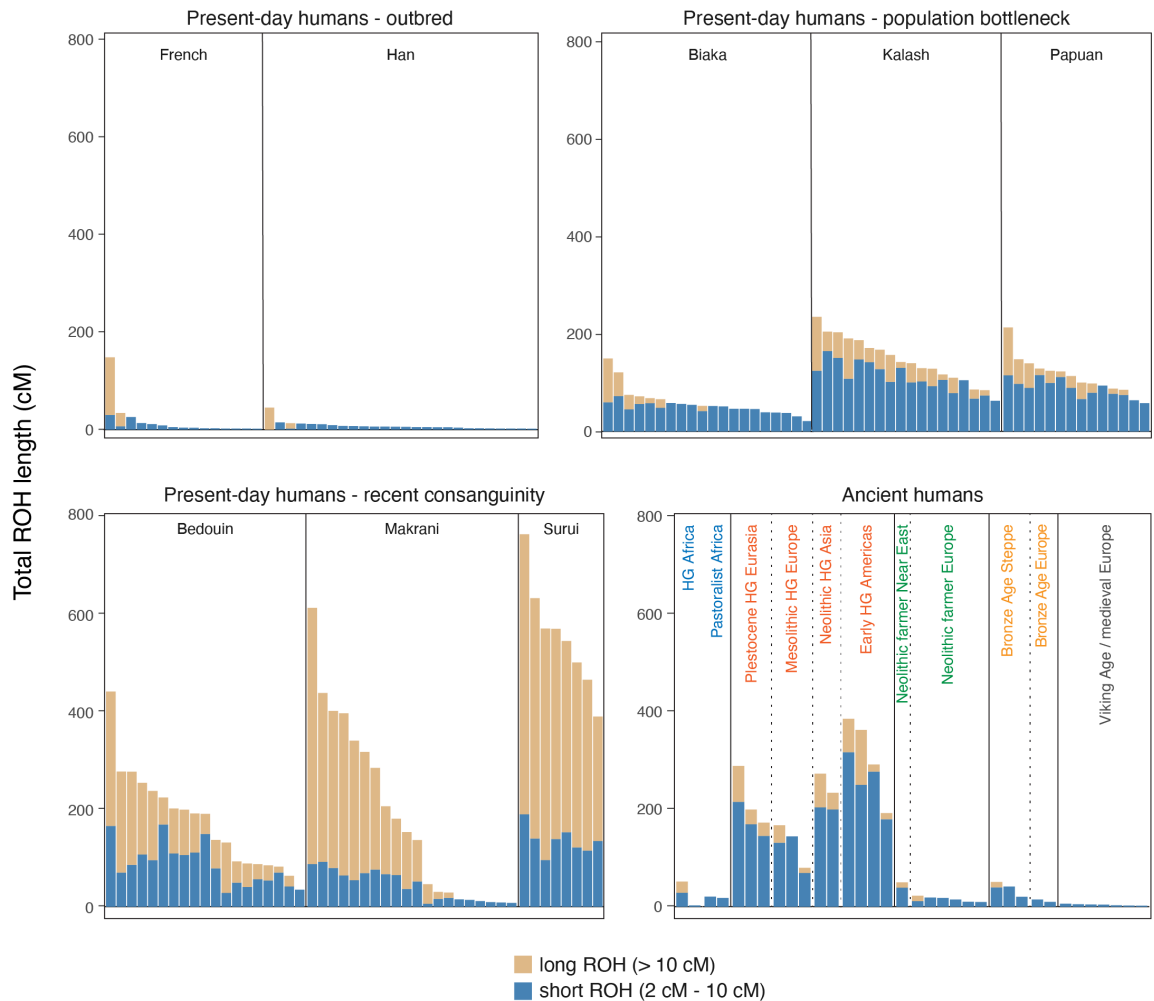

**Figure 3. Homozygosity-by-descent in present-day and ancient humans.** Barplots showing the total length of the genome contained in runs of homozygosity (ROH) for selected present-day and ancient individuals. For each individual, total ROH lengths are further stratified into short (blue) or long (beige) segments, reflecting signatures of different demographic processes. Outbreeding present-day populations with large effective population sizes (French Europeans, Han Chinese) are characterized by few short ROHs; present-day populations with smaller effective population sizes (Biaka, Kalash, Papuans) show increased total ROH length from large numbers of short ROHs; populations with recent consanguinity (Bedouins, Makrani, Surui) show the highest length of ROH, a large fraction of which is contained in long segments (> 10 cM). Ancient individuals show marked differences in their ROH length distributions, corresponding to their age and modes of subsistence. Ancient hunter-gatherers (HG) from Eurasia and the Americas show high total ROH lengths mostly contained in shorter segments, consistent with small effective population sizes without evidence of recent consanguinity[89,93]. Later groups from farming and pastoralist societies have markedly reduced ROH levels, similar to large outbred present-day populations. ROH were inferred using *ibdSeq*[82] on a dataset of publicly available SNP genotype data of present-day humans[14] combined with diploid genotypes for ancient humans with median genomic coverage > 8X (Supplementary Table 1).

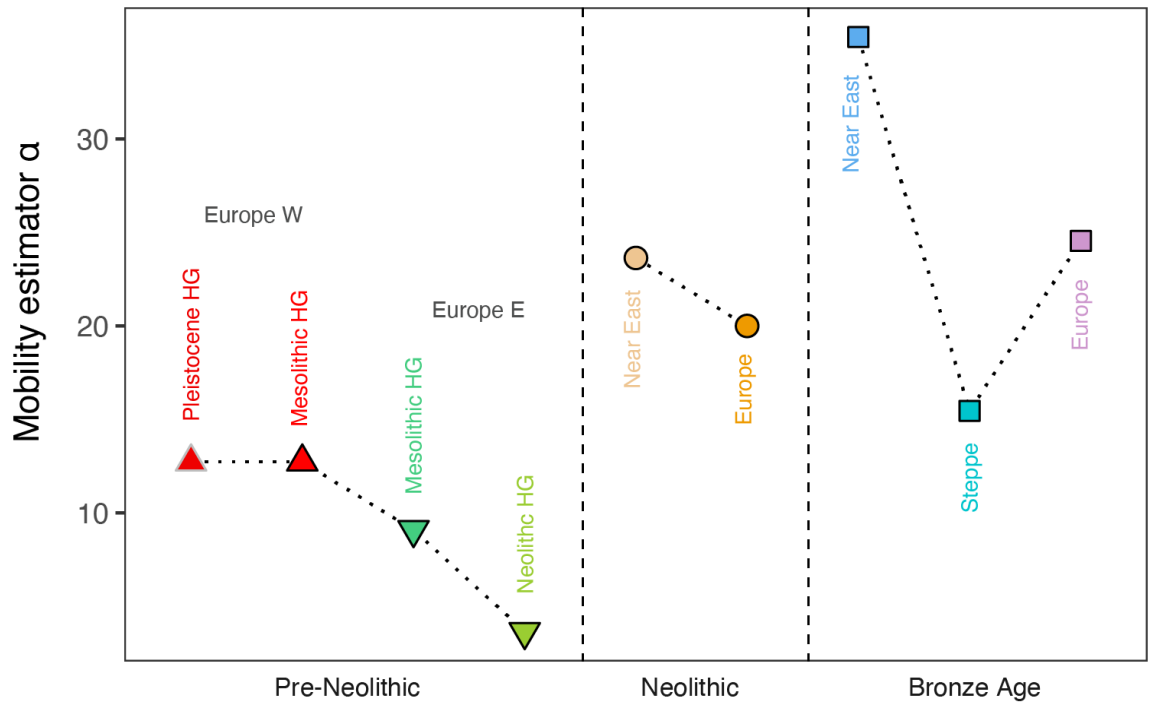

**Figure 4. Changes in patterns of mobility through time.** We applied the Loog et al.[55] mobility estimator to a dataset compiled of 774 publicly available ancient genomes from Western Eurasia across various periods of time (HG: hunter-gatherers). Genetic distances between pairs of individuals were estimated as 1-p(IBS), where p(IBS) is the fraction of alleles shared identical-by-state between the individuals. A full list of samples with group labels, geographic location and median age is provided in Supplementary Table 2.

**Glossary**

Admixture: the introduction of genetic lineages from one population into another population that is genetically differentiated from it, because of interbreeding between them at some point in the past.

Agent-based model: computational model designed for simulating the behavior of multiple autonomous agents that may interact with each other, so as to study their collective effects on a system.

Demic diffusion: the spread of technologies or cultures via movement of people.

Exogamy: the cultural practice by which females or males tend to marry outside their immediate kin group.

Identical by descent (IBD): two segments from two different genomes are IBD if they were both inherited from a recent ancestor shared between the two genomes.

Homozygous by descent (HBD): Genomic segments shared IBD within the same individual; resulting in continuous stretches of homozygous genotypes termed runs of homozygosity (ROH).

Kriging: a geostatistical method of interpolation on a spatial grid, by which unknown values are inferred via a Gaussian process model from known (but often sparsely and unevenly sampled) values. It was developed by Danie Krige and Georges Matheron in the 1960s.

Megalith: large stone structure, tomb or monument. In Europe, the practice of megalith construction mainly took off in the Neolithic period, reached an apogee during the Chalcolithic period and continued into the Bronze Age.

Patrilineality: kinship system in which a person's social status, family membership and/or property rights are determined through that person's paternal lineage. In contrast, in a matrilineal system, these are determined through the maternal lineage.

Isotope analysis: the study of the concentrations of different varieties of a chemical element - like carbon, nitrogen or strontium - that have different numbers of neutrons in biological samples. They can indicate the relative abundance of vegetation types, dietary items in archaeological sites or identify non-local individuals.

Uniparental markers: sequences of DNA that are - barring rare exceptions - inherited only from one or another of a person's parents. Examples include the mitochondrial DNA genome (transmitted from the mother alone) and the Y-chromosome genome (transmitted from fathers to sons).